\documentclass[aps,prl,twocolumn,groupedaddress,showpacs,floatfix]{revtex4}

\usepackage{graphicx} 
\usepackage{mathptmx} 
\usepackage{amssymb} 
\begin{document}

\title{Passive Cooling of a Micromechanical Oscillator with a Resonant
  Electric Circuit}

\author{K. R. Brown}
\altaffiliation{krbrown@boulder.nist.gov}
\author{J. Britton}
\author{R. J. Epstein}
\author{J. Chiaverini}
\altaffiliation{Present address: Los Alamos National Laboratory, MS 
D454, P-21, Los Alamos, NM 87545, USA.}
\author{D. Leibfried}
\author{D. J. Wineland}
\affiliation{Time and Frequency Division, National Institute of
  Standards and Technology, 325 Broadway, Boulder, CO 80305, USA}

\date{8 May 2007}

\begin{abstract}
  We cool the fundamental mode of a miniature cantilever by
  capacitively coupling it to a driven rf resonant circuit. Cooling
  results from the rf capacitive force, which is phase shifted
  relative to the cantilever motion. We demonstrate the technique by
  cooling a 7~kHz cantilever from room temperature to 45~K, obtaining
  reasonable agreement with a model for the cooling, damping, and
  frequency shift. Extending the method to higher frequencies in a
  cryogenic system could enable ground state cooling and may prove
  simpler than related optical experiments in a low temperature
  apparatus.
\end{abstract}


\maketitle

Stimulated by the early work of Braginsky and collaborators
\cite{braginsky70,braginsky77}, the quantum-limited measurement and
control of mechanical oscillators continues to be a subject of great
interest. If one can cool to the ground state of the oscillator, the
generation of nonclassical states of motion also becomes feasible. For
an atom bound in a harmonic well, laser cooling in a room-temperature
apparatus can cool the modes of mechanical motion to a level with mean
occupation numbers $\langle n \rangle < 0.1$ for oscillation
frequencies $\sim$1$-$10 MHz \cite{diedrich89,monroe_cooling}. This
has made it possible to generate nonclassical mechanical oscillator
states such as squeezed, Fock \cite{meekhof96}, multiparticle
entangled \cite{monroe96_cats}, and (in principle) arbitrary
superposition states \cite{ben-kish03}.

For more macroscopic systems, smaller and smaller micromechanical
resonators have approached the quantum limit through thermal contact
with a cryogenic bath (for a summary, see \cite{schwab05}). Small
mechanical resonators, having low-order mode frequencies of 10$-$1000
MHz, can come close to the quantum regime at low temperature ( $<
1$~K), and mean occupation numbers of approximately 25 have been
achieved \cite{naik06}. Cooling of macroscopic mechanical oscillators
also has been achieved with optical forces. The requisite damping can
be implemented by use of active external electronics to control the
applied force \cite{cohadon99,arcizet_sep06,weld06,kleckner06} (see
also \cite{poggio07}). Passive feedback cooling has been realized in
which a mirror attached to a mechanical oscillator forms an optical
cavity with another stationary mirror. For appropriate tuning of
radiation incident on the cavity, a delay in the optical force on the
oscillator as it moves gives cooling. This delay can result from a
photothermal effect \cite{metzger04,harris07} or from the stored
energy response time of the cavity
\cite{arcizet_nov06,gigan06,corbitt07}. Closely related passive
cooling has been reported in \cite{schliesser06,naik06}.

We demonstrate a similar cooling mechanism where the damping force is
the electric force between capacitor plates \cite{milatz53} that here
contribute to a resonant rf circuit
\cite{braginsky77,wineland06}. This approach has potential practical
advantages over optical schemes: eliminating optical components
simplifies fabrication and integration into a cryogenic system, and
the rf circuit could be incorporated on-chip with the mechanical
oscillator.

\begin{figure}
\centering \includegraphics{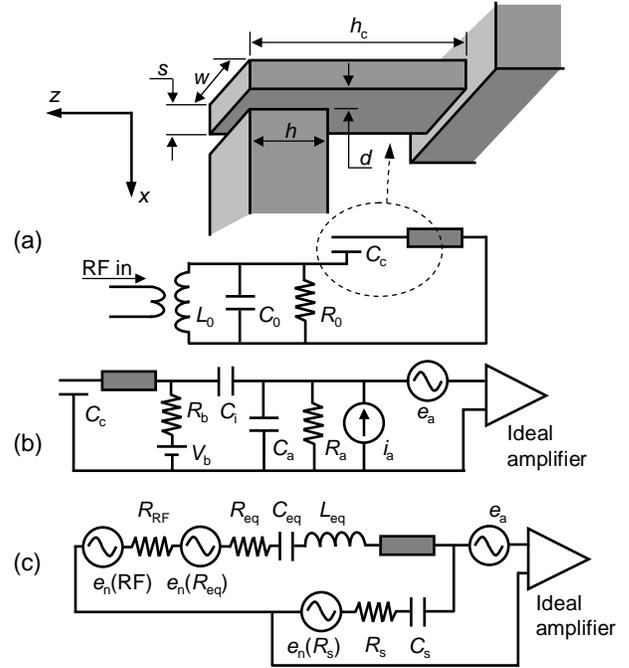}
\caption{Schematics of the cantilever cooling and detection
  electronics. (a) Cantilever and associated rf circuitry. (b)
  Motional detection electronics. Near $\omega_c$ the rf
  circuit looks like a short to ground as shown. (c) Equivalent
  circuit for the cantilever and detection electronics near $\omega
  \approx \omega_c$.}
\label{fig:schematic}
\end{figure}

A conducting cantilever of mass density $\rho$ is fixed at one end
[Fig.~\ref{fig:schematic}(a)].
One face is placed a distance $d$ from a rigidly mounted plate of area
$w \times h$, forming a parallel-plate capacitor $C_c =
\epsilon_0wh/d$, where $\epsilon_0$ is the vacuum dielectric constant.
An inductor $L_0$ and capacitor $C_0$ in parallel with $C_c$ form a
resonant rf circuit with frequency $\Omega_0 = 1/\sqrt{L_0(C_0 +
  C_c)}$ and with losses represented by resistance $R_0$. We assume
$Q_\mathrm{rf} \gg 1$, where $Q_\mathrm{rf} = \Omega_0 L_0/R_0 =
\Omega_0/\gamma$ and $\gamma$ is the damping rate.

We consider the lowest-order flexural mode of the cantilever, where
the free end oscillates in the $\hat{x}$ direction [vertical in
Fig.~\ref{fig:schematic}(a)] with angular frequency $\omega_c \ll
\gamma$. We take $x$ to be the displacement at the end of the
cantilever, so the displacement as a function of (horizontal) position
$z$ along the length of the cantilever is given by $x(z) = f(z)x$,
where $f(z)$ is the mode function (see, e.g., \cite{butt95}). Small
displacements due to a force $F$ can be described by the equation of
motion,
\begin{equation}
m\ddot{x} + m \Gamma \dot{x} + m \omega_c^2 x = F,
\label{equation_of_motion}
\end{equation}
where $\Gamma$ is the cantilever damping rate and $m$ its effective
mass, given by $\rho \xi_c'' w h_c s$, where
$\xi_c'' \equiv \frac{1}{h_c} \int_{h_c}
f(z)^2 \, dz = 0.250$ for a rectangular beam.

For simplicity, first assume $h \ll h_c$, so that the force
is concentrated at the end of the cantilever. If a potential $V$ is
applied across $C_c$, the capacitor plates experience a
mutual attractive force $F = \epsilon_0 w h V^2/(2 d^2) = C_c
V^2/(2 d)$.  Consider that $V$ is an applied rf potential
$V_\mathrm{rf} \cos (\Omega_\mathrm{rf} t)$ with $\Omega_\mathrm{rf}
\approx \Omega_0$. Because $\omega_c \ll \Omega_0$, the force
for frequencies near $\omega_c$ can be approximated by the
time-averaged rf force
\begin{equation}
  F_\mathrm{rf} = \frac{C_c \langle V^2 \rangle}{2 d} = 
  \frac{C_c V_\mathrm{rf}^2}{4 d} = \frac{\epsilon_0
  w h V_\mathrm{rf}^2}{4 d^2},
\label{capacitor_force}
\end{equation}
where, for a fixed input rf power, $V_\mathrm{rf}$ will depend on
$\Delta \Omega \equiv \Omega_0 - \Omega_{\mathrm{rf}}$, according to
\begin{equation}
\frac{V_\mathrm{rf}^2}{V_\mathrm{max}^2} = \frac{1}{1 + \bigl[2
Q_{\mathrm{rf}} \Delta \Omega/\Omega_0 \bigr]^2} \equiv
\mathcal{L}(\Delta \Omega)\ .
\label{eq:L}
\end{equation}
As the cantilever oscillates, its motion modulates the capacitance of
the rf circuit
thereby modulating
$\Omega_0$. As $\Omega_0$ is modulated relative to
$\Omega_{\mathrm{rf}}$, so too is the rf potential across the
capacitance, according to Eq.~(\ref{eq:L}). The associated modulated
force shifts the cantilever's resonant frequency. Because of the finite
response time of the rf circuit, there is a phase lag in the force
relative to the motion.  For $\Delta \Omega > 0$ the phase lag leads
to a force component that opposes the cantilever velocity, leading to
damping. If this damping is achieved without adding too much force
noise then it cools the cantilever.

The average force due to applied potentials displaces the equilibrium
position $d_0$ of the cantilever. We assume this displacement is small
and is absorbed into the definition of $d_0$, writing \cite{d0x} $d
\equiv d_0 - x$, where $x \ll d_0$.~
Following \cite{braginsky77} or \cite{wineland06} we find
$\omega_c^2 \rightarrow \omega_c^2 (1 - \kappa)$ and
$\Gamma \rightarrow \Gamma + \Gamma'$, with
%
\begin{eqnarray}
  \kappa & \equiv & \frac{C_c V_\mathrm{max}^2 \mathcal{L}(\Delta \Omega)}
  {2 m \omega_c^2 d_0^2} \biggl[ \xi'' + \frac{2 (\xi')^2 Q_{\mathrm{rf}} 
    \Delta \Omega \mathcal{L}(\Delta \Omega)}{\gamma} \nonumber \\
  & & \times \frac{C_c}{C_c + C_0} \biggr],
\label{kappa}
\end{eqnarray}
\begin{equation}
\Gamma' \equiv  \frac{Q_{\mathrm{rf}} V_\mathrm{max}^2 C_c^2}{m
\omega_c d_0^2 (C_c + C_0)} \frac{(\xi')^2 \Delta
\Omega \mathcal{L}(\Delta\Omega)^2}{\gamma} \sin{\phi},
\label{gammaprime}
\end{equation}
where $\xi' \equiv \frac{1}{h} \int_h f(z)\,dz$ and $\xi''
\equiv \frac{1}{h} \int_h f(z)^2\,dz$ are geometrical factors
required when $h \ll h_c$ is not satisfied.  The phase $\phi$
is equal to $\omega_c \tau$, where $\tau =
4\mathcal{L}(\Delta \Omega)/\gamma$ is the response time of the rf
circuit \cite{marquardt07}. For $\Delta \Omega > 0$, $\Gamma'$ gives
increased damping.  For $\Delta \Omega = \gamma/2$ and $h \ll
h_c$ ($\xi' \approx \xi'' \approx 1)$, we obtain the
expressions of \cite{wineland06}.  For our experiment, $h \approx
h_c$, $\xi' = 0.392$, and $\xi'' = \xi_c'' = 0.250$.

We detect the cantilever's motion by biasing it with a static
potential $V_b$ through resistor $R_b$ as shown in
Fig.~\ref{fig:schematic}(b), where $R_a$, $C_a$,
$i_a$, and $e_a$ represent the equivalent input
resistance, capacitance, current noise, and voltage noise,
respectively, of the detection amplifier. We make $R_a$ and
$R_b$ large to minimize their contribution to the current
noise $i_a$. We assume $C_i \gg (C_c +
C_a)$ and $\omega_c R (C_c + C_a)
\gg 1$, where $1/R \equiv 1/R_b + 1/R_a$. As the
cantilever moves, thereby changing $C_c$, it creates a
varying potential that is detected by the amplifier.

The (charged) cantilever can be represented by the series electrical
circuit in Fig.~\ref{fig:schematic}(c). From
Eq.~(\ref{capacitor_force}) and following \cite{wineland75}, the
equivalent inductance is given by $L_\mathrm{eq} = m
d_0^2/(q_c\xi')^2$, where $q_c$ is the average
charge on the cantilever. From $L_\mathrm{eq}$, $\omega_c$,
and $\Gamma$, we can then determine $C_\mathrm{eq} =
1/(\omega_c^2 L_\mathrm{eq})$ and $R_\mathrm{eq} =
L_\mathrm{eq} \Gamma$. Additional damping due to the rf force is
represented by $R_\mathrm{rf} = L_\mathrm{eq} \Gamma'$. For
frequencies $\omega \approx \omega_c$, the parallel
combination of $R_b$, $C_a$, and $R_a$ can
also be expressed instead as the Th\'{e}venin equivalent
$R_s$-$C_s$ circuit in
Fig.~\ref{fig:schematic}(c). The amplifier's current noise
$i_a$ is now represented as $e_n(R_s)$. The
intrinsic thermal noise of the cantilever is characterized by a noise
voltage $e_n(R_\mathrm{eq})$ having spectral density $4
k_B T_c R_\mathrm{eq}$, where $k_B$ is
Boltzmann's constant and $T_c$ is the cantilever temperature.

We must also consider noise from the rf circuit. In
Eq.~(\ref{capacitor_force}), we replace $V_\mathrm{rf}$ with
$V_\mathrm{rf}+ v_n(\mathrm{rf})$, where
$v_n(\mathrm{rf})$ is the noise across the cantilever
capacitance $C_c$ due to resistance in the rf circuit and
noise injected from the rf source. The cantilever is affected by
amplitude noise $v_n(\mathrm{rf})$ at frequencies near
$\Omega_{\mathrm{rf}} \pm \omega_c$, because cross terms in
Eq.~(\ref{capacitor_force}) give rise to random forces at the
cantilever frequency.  This force noise can be represented by
$e_n(\mathrm{rf})$ in the equivalent circuit. The noise terms
sum to $e_n^2 = e_n^2(R_\mathrm{eq}) +
e_n^2(R_s) + e_n^2(\mathrm{rf})$
($e_a$ does not drive the cantilever), which gives a
cantilever effective temperature
\begin{equation}
T_\mathrm{eff}=\frac{e_n^2
  }{4k_B(R_\mathrm{eq} + R_\mathrm{rf} + R_s)}.
\label{effective_temperature}
\end{equation}

Our cantilever has nominal dimensions $h_ c \approx 1.5$~mm,
$s \approx 14$~$\mu$m \cite{uncertainties}, and $w \approx
200$~$\mu$m, created by etching through a $p^{++}$-doped ( $\sim
0.001$ $\Omega$ cm), 200~$\mu$m thick silicon wafer with a standard
Bosch reactive-ion-etching process. Its resonant frequency and quality
factor are $\omega_c/(2\pi) \approx 7$~kHz and $Q \approx
20\:000$. The cantilever is separated by $d_0 \approx 16$~$\mu$m
\cite{uncertainties} from a nearby doped silicon rf electrode, forming
capacitance $C_c$. The sample is enclosed in a vacuum chamber
with pressure less than $10^{-5}$~Pa. The rf electrode is connected
via a vacuum feedthrough to a quarter-wave resonant cavity with $L_0 =
330(30)$~nH and with loaded quality factor $Q_\mathrm{rf}=234(8)$ at
$\Omega_\mathrm{rf}/(2 \pi)=100$~MHz when impedance matched to the
source. The cantilever is connected by a short length of coaxial cable
and blocking capacitor $C_i=4$~nF to a low-noise junction
field-effect transistor amplifier [see
Fig.~\ref{fig:schematic}(b)]. We have $C_a=48(1)$~pF, with
$R_a=R_b=1$~G$\Omega$. We use $V_b =
-50$~V, which gives a measured 2.5~$\mu$m static deflection at the
cantilever end.

We temporarily lowered $R_a$ to 600~k$\Omega \approx
1/(\omega_cC_a)$, in which case the cantilever noise
spectrum strongly distorts from a Lorentzian lineshape (not shown),
and it becomes straightforward to extract the equivalent circuit
parameters of Fig.~\ref{fig:schematic}(c). We find $L_\mathrm{eq} =
27\:000(600)$~H. To lowest order in rf power this equivalent inductance
remains constant, so we assume this value for $L_\mathrm{eq}$ in
subsequent fits to the thermal spectra, while $R_\mathrm{rf}$ is
allowed to vary to account for rf power induced changes in the
cantilever damping.

\begin{figure}
\centering \includegraphics{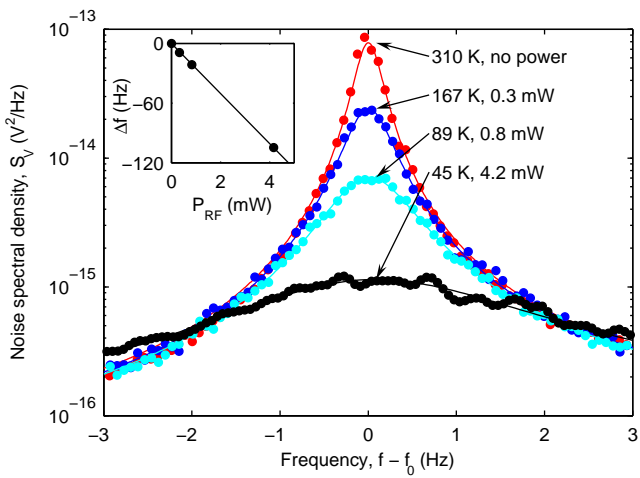}
\caption{(color online). Cantilever thermal spectra for four values of
  rf power. The $x$ axis for each spectrum has been shifted to align
  the maxima of the three data sets. $S_v$ is the measured
  noise referred to the input of the amplifier. Solid lines are fits
  to the model in Fig.~\ref{fig:schematic}(c), giving the temperatures
  indicated by arrows. (Inset) Cantilever frequency shift $\Delta f$
  vs rf power.}
\label{fig:thermal_spectra}
\end{figure}

For $R_a = 1$~G$\Omega$ we measure
$e_a=1.5$~nV/$\sqrt{\mathrm{Hz}}$ and
$i_a=16$~fA/$\sqrt{\mathrm{Hz}}$.
Figure~\ref{fig:thermal_spectra} shows a series of thermal spectra
acquired with this value of $R_a$ at different values of rf
power $P_\mathrm{rf}$ but at constant detuning $\Delta\Omega = 2 \pi
\times 90$~kHz $= 0.21 \gamma$.
Both the lowering and the broadening of the spectra with increasing
$P_\mathrm{rf}$ are evident, in accordance with
Eq.~(\ref{gammaprime}). Here, the effective temperature is very nearly
proportional to the area under the curves, although there is a slight
asymmetric distortion from a Lorentzian line shape, fully accounted for
by the equivalent circuit model. The center frequency of each spectrum
also shifts to lower frequencies for increasing $P_\mathrm{rf}$, as
predicted by Eq.~(\ref{kappa}) and the definition of $\kappa$ in terms
of $\omega_c$. After calibrating the gain of the amplifier,
we extract $e_n^2$ for each spectrum from a fit to the model
of Fig.~\ref{fig:schematic}(c). The absolute effective temperature is
then given by Eq.~(\ref{effective_temperature}).

\begin{figure}
\centering \includegraphics{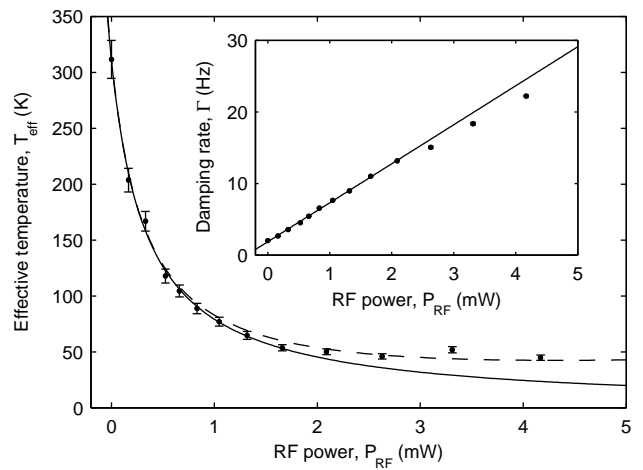}
\caption{$T_\mathrm{eff}$ as a function of rf power. The solid line is
  the temperature predicted by Eq.~(\ref{effective_temperature})
  using $\Gamma$ from the fit in the inset and $e_n =
  e_n(R_\mathrm{eq})$, while the dashed line takes into
  account additional noise due to the rf source. (Inset) Damping rate
  vs rf power. The solid line is a linear fit to the first ten
  points.}
\label{fig:effective temperature}
\end{figure}

Equations~(\ref{gammaprime}) and (\ref{effective_temperature}) predict
that the cantilever's effective temperature should fall with
increasing $P_\mathrm{rf}$, as demonstrated by the data in
Fig.~\ref{fig:effective temperature} for low power.
With no rf applied we find $T_\mathrm{eff} = 310(20)$~K. The coldest
spectrum corresponds to a temperature of 45(2)~K, a factor of 6.9
reduction. The minimum temperature appears to be limited by AM noise
from our rf source. This noise power at $\Omega_\mathrm{rf} \pm
\omega_c$ is constant relative to the carrier, leading to a
noise power at $\omega_c$ given by
$e_n(\mathrm{rf})^2 \propto P_\mathrm{rf}^2$. We fit the
residual noise $e_n(\mathrm{rf})^2$ to a quadratic in
$P_\mathrm{rf}$, giving the dashed line temperature prediction in
Fig.~\ref{fig:effective temperature}. From this fit we determine that
the AM noise of our source is -170~dBc/Hz, reasonably consistent with
the value (-167~dBc/Hz) measured by spectrum analysis.

The inset shows the cantilever damping rate $\Gamma$ versus
$P_\mathrm{rf}$. The slope is $\Gamma'/P_\mathrm{rf} =
5450(70)$~Hz/W, slightly higher than the value 3970~Hz/W calculated
from Eq.~(\ref{gammaprime}) and the nominal cantilever dimensions. The
nonlinearity in $\Gamma'/P_\mathrm{rf}$ at higher powers is consistent
with the cantilever being pulled toward the rf electrode. We have
numerically simulated this effect and find reasonable agreement. The
variation of $\kappa$ with $P_\mathrm{rf}$ (not shown) is also linear,
with a slope $\kappa/P_\mathrm{rf} = 7.64(8)$~$\mathrm{W}^{-1}$,
compared with the value 3.45~$\mathrm{W}^{-1}$ calculated from
Eq.~(\ref{kappa}).

Although $\Gamma'/P_\mathrm{rf}$ and $\kappa/P_\mathrm{rf}$ differ
from their predicted values, this disagreement is not unexpected
considering the relatively large variations in dimensions $d_0$ and
$s$ \cite{uncertainties}. Another indication of these uncertainties is
that optical measurements of the static deflection of the cantilever
along its length disagree with predictions based on a constant
cantilever cross section.  This will lead to deviations from our
calculated values of $\xi'$, $\xi''$, and $\xi_c''$.
However, we stress that these deviations should not give significant
errors in our measured values of $L_\mathrm{eq}$, $R_\mathrm{eq}$, and
therefore our determination of $T_\mathrm{eff}$.

\begin{figure}
\centering \includegraphics{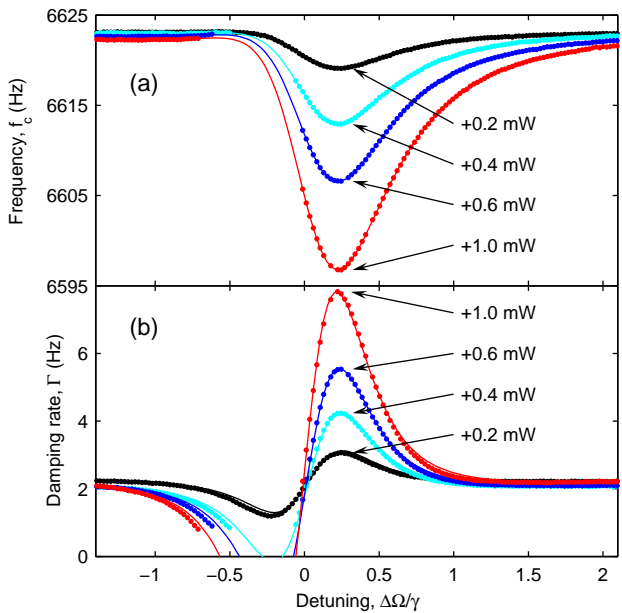}
\caption{(color online). Variation of the cantilever resonance with
  respect to rf frequency. (a) Cantilever frequency
  $f_c$ and (b) damping rate $\Gamma$ vs
  normalized rf detuning $\Delta\Omega/\gamma$ for several values of
  $P_\mathrm{rf}$. The missing points between $-$0.7~MHz and 0~MHz
  correspond to a region of instability where $\Gamma$ becomes
  negative. Solid lines are fits to Eqs.~(\ref{kappa}) and
  (\ref{gammaprime}).}
\label{fig:ringdown}
\end{figure}

To further test the model, we examine $\Gamma$ and $\omega_c$
as a function of $\Delta\Omega$ (Fig.~\ref{fig:ringdown}).
For large detunings $\Delta\Omega$, $\Gamma$ asymptotically approaches
the value obtained in Fig.~\ref{fig:effective temperature} for
$P_\mathrm{rf}=0$. Near $\Delta\Omega=0$, $f_c$ is generally
shifted to a lower value, while $\Gamma$ is either enhanced or
suppressed, according to the sign of $\Delta \Omega$. Data cannot be
obtained for $\Delta\Omega < 0$ when the rf power level is sufficient
to drive the cantilever into instability ($\Gamma<0$). The solid line
fits show good agreement with the predicted behavior. From these fits
we extract $C_c = 0.09$~pF, lower than the value $0.17$~pF
obtained from the physical dimensions. This disagreement is not
surprising for the reasons mentioned above.

Some experiments using optical forces have observed strong effects
from the laser power absorbed in the cantilever mirror. A conservative
estimate of the rf power dissipated in our cantilever gives a
temperature rise of less than 1~K at the highest power, so these
effects should not be significant.

Although rather modest cooling is obtained here, the basic method
could eventually provide ground state cooling. For this we must
achieve the resolved sideband limit, where $\omega_c >
\gamma$ \cite{marquardt07,wilson-rae07}, and the cooling would be very
similar to the atomic case \cite{diedrich89,monroe_cooling}. To insure
a mean quantum number $n$ less than one, the heating rate from the
ground state $\dot{n}_\mathrm{heat} = \Gamma k_B
T_c/(\hbar \omega_c)$ must be less than the cooling
rate for $n = 1 \rightarrow 0$. The cooling rate
$\dot{n}_\mathrm{cool}$ can be estimated by noting that each absorbed
photon on the lower sideband (at the applied rf frequency $\Omega_0 -
\omega_c$) is accompanied by reradiation on the rf
``carrier'' at $\Omega_0$. If we assume the lower sideband is
saturated for $n \approx 1$, $\dot{n}_\mathrm{cool} \approx
\gamma/2$. Hence we require $R \equiv
\dot{n}_\mathrm{heat}/\dot{n}_\mathrm{cool} \approx 2 k_B
T_c Q_\mathrm{rf}/(\hbar \Omega_0 Q_c) \ll 1$. For
example, if $T_c = 0.1$~K, $\Omega_0/(2 \pi) = 20$~GHz,
$Q_\mathrm{rf} = 5000$ (e.g., a stripline), and $Q_c =
20\:000$ we have $R \approx 0.05$. For resolved sidebands, we require
$\omega_c/(2 \pi) > 4$~MHz.

\begin{acknowledgments}
  We thank S. Voran for the use of his lab space. We thank D. Howe,
  C. Nelson, and A. Hati for providing us with low-noise rf sources
  and for discussion, and we thank R. Simmonds and J. Bollinger
  for comments. This Letter, a submission of NIST, is not subject to
  US copyright.
\end{acknowledgments}

\bibliography{dave_w_et_al_style,kenton_et_al_style}

\begin{thebibliography}{28}
\expandafter\ifx\csname natexlab\endcsname\relax\def\natexlab#1{#1}\fi
\expandafter\ifx\csname bibnamefont\endcsname\relax
  \def\bibnamefont#1{#1}\fi
\expandafter\ifx\csname bibfnamefont\endcsname\relax
  \def\bibfnamefont#1{#1}\fi
\expandafter\ifx\csname citenamefont\endcsname\relax
  \def\citenamefont#1{#1}\fi
\expandafter\ifx\csname url\endcsname\relax
  \def\url#1{\texttt{#1}}\fi
\expandafter\ifx\csname urlprefix\endcsname\relax\def\urlprefix{URL }\fi
\providecommand{\bibinfo}[2]{#2}
\providecommand{\eprint}[2][]{\url{#2}}

\bibitem[{\citenamefont{Braginski\u{i}
  et~al.}(1970)\citenamefont{Braginski\u{i}, Manukin, and {Yu.
  Tikhonov}}}]{braginsky70}
\bibinfo{author}{\bibfnamefont{V.~B.} \bibnamefont{Braginski\u{i}}},
  \bibinfo{author}{\bibfnamefont{A.~B.} \bibnamefont{Manukin}},
  \bibnamefont{and} \bibinfo{author}{\bibfnamefont{M.}~\bibnamefont{{Yu.
  Tikhonov}}}, \bibinfo{journal}{Sov. Phys. JETP}
  \textbf{\bibinfo{volume}{31}}, \bibinfo{pages}{829} (\bibinfo{year}{1970}).

\bibitem[{\citenamefont{Braginsky and Manukin}(1977)}]{braginsky77}
\bibinfo{author}{\bibfnamefont{V.~B.} \bibnamefont{Braginsky}}
  \bibnamefont{and} \bibinfo{author}{\bibfnamefont{A.~B.}
  \bibnamefont{Manukin}}, \emph{\bibinfo{title}{Measurement of Weak Forces in
  Physics Experiments}} (\bibinfo{publisher}{The University of Chicago Press},
  \bibinfo{address}{Chicago}, \bibinfo{year}{1977}).

\bibitem[{\citenamefont{Diedrich et~al.}(1989)\citenamefont{Diedrich,
  Bergquist, Itano, and Wineland}}]{diedrich89}
\bibinfo{author}{\bibfnamefont{F.}~\bibnamefont{Diedrich}},
  \bibinfo{author}{\bibfnamefont{J.~C.} \bibnamefont{Bergquist}},
  \bibinfo{author}{\bibfnamefont{W.~M.} \bibnamefont{Itano}}, \bibnamefont{and}
  \bibinfo{author}{\bibfnamefont{D.~J.} \bibnamefont{Wineland}},
  \bibinfo{journal}{Phys. Rev. Lett.} \textbf{\bibinfo{volume}{62}},
  \bibinfo{pages}{403} (\bibinfo{year}{1989}).

\bibitem[{\citenamefont{Monroe~\emph{et al}.}(1995)}]{monroe_cooling}
\bibinfo{author}{\bibfnamefont{C.}~\bibnamefont{Monroe~\emph{et al}.}},
  \bibinfo{journal}{Phys. Rev. Lett.} \textbf{\bibinfo{volume}{75}},
  \bibinfo{pages}{4011} (\bibinfo{year}{1995}).

\bibitem[{\citenamefont{Meekhof~\emph{et al}.}(1996)}]{meekhof96}
\bibinfo{author}{\bibfnamefont{D.~M.} \bibnamefont{Meekhof~\emph{et al}.}},
  \bibinfo{journal}{Phys. Rev. Lett.} \textbf{\bibinfo{volume}{76}},
  \bibinfo{pages}{1796} (\bibinfo{year}{1996}).

\bibitem[{\citenamefont{Monroe et~al.}(1996)\citenamefont{Monroe, Meekhof,
  King, and Wineland}}]{monroe96_cats}
\bibinfo{author}{\bibfnamefont{C.}~\bibnamefont{Monroe}},
  \bibinfo{author}{\bibfnamefont{D.~M.} \bibnamefont{Meekhof}},
  \bibinfo{author}{\bibfnamefont{B.~E.} \bibnamefont{King}}, \bibnamefont{and}
  \bibinfo{author}{\bibfnamefont{D.~J.} \bibnamefont{Wineland}},
  \bibinfo{journal}{Science} \textbf{\bibinfo{volume}{272}},
  \bibinfo{pages}{1131} (\bibinfo{year}{1996}).

\bibitem[{\citenamefont{Ben-{K}ish \emph{et al.}}(2003)}]{ben-kish03}
\bibinfo{author}{\bibfnamefont{A.}~\bibnamefont{Ben-{K}ish \emph{et al.}}},
  \bibinfo{journal}{Phys. Rev. Lett.} \textbf{\bibinfo{volume}{90}},
  \bibinfo{pages}{037902} (\bibinfo{year}{2003}).

\bibitem[{\citenamefont{Schwab and Roukes}(2005)}]{schwab05}
\bibinfo{author}{\bibfnamefont{K.~C.} \bibnamefont{Schwab}} \bibnamefont{and}
  \bibinfo{author}{\bibfnamefont{M.~L.} \bibnamefont{Roukes}},
  \bibinfo{journal}{Phys. Today} \textbf{\bibinfo{volume}{58}},
  \bibinfo{pages}{No. 7, 36} (\bibinfo{year}{2005}).

\bibitem[{\citenamefont{Naik~\emph{et al}.}(2006)}]{naik06}
\bibinfo{author}{\bibfnamefont{A.}~\bibnamefont{Naik~\emph{et al}.}},
  \bibinfo{journal}{Nature (London)} \textbf{\bibinfo{volume}{443}},
  \bibinfo{pages}{193} (\bibinfo{year}{2006}).

\bibitem[{\citenamefont{Cohadon et~al.}(1999)\citenamefont{Cohadon, Heidmann,
  and Pinard}}]{cohadon99}
\bibinfo{author}{\bibfnamefont{P.-F.} \bibnamefont{Cohadon}},
  \bibinfo{author}{\bibfnamefont{A.}~\bibnamefont{Heidmann}}, \bibnamefont{and}
  \bibinfo{author}{\bibfnamefont{M.}~\bibnamefont{Pinard}},
  \bibinfo{journal}{Phys. Rev. Lett.} \textbf{\bibinfo{volume}{83}},
  \bibinfo{pages}{3174} (\bibinfo{year}{1999}).

\bibitem[{\citenamefont{Arcizet~\emph{et
  al}.}(2006{\natexlab{a}})}]{arcizet_sep06}
\bibinfo{author}{\bibfnamefont{O.}~\bibnamefont{Arcizet~\emph{et al}.}},
  \bibinfo{journal}{Phys. Rev. Lett.} \textbf{\bibinfo{volume}{97}},
  \bibinfo{pages}{133601} (\bibinfo{year}{2006}{\natexlab{a}}).

\bibitem[{\citenamefont{Weld and Kapitulnik}(2006)}]{weld06}
\bibinfo{author}{\bibfnamefont{D.~M.} \bibnamefont{Weld}} \bibnamefont{and}
  \bibinfo{author}{\bibfnamefont{A.}~\bibnamefont{Kapitulnik}},
  \bibinfo{journal}{Appl. Phys. Lett.} \textbf{\bibinfo{volume}{89}},
  \bibinfo{pages}{164102} (\bibinfo{year}{2006}).

\bibitem[{\citenamefont{Kleckner and Bouwmeester}(2006)}]{kleckner06}
\bibinfo{author}{\bibfnamefont{D.}~\bibnamefont{Kleckner}} \bibnamefont{and}
  \bibinfo{author}{\bibfnamefont{D.}~\bibnamefont{Bouwmeester}},
  \bibinfo{journal}{Nature (London)} \textbf{\bibinfo{volume}{444}},
  \bibinfo{pages}{75} (\bibinfo{year}{2006}).

\bibitem[{\citenamefont{Poggio et~al.}(2007)\citenamefont{Poggio, Degen, Mamin,
  and Rugar}}]{poggio07}
\bibinfo{author}{\bibfnamefont{M.}~\bibnamefont{Poggio}},
  \bibinfo{author}{\bibfnamefont{C.~L.} \bibnamefont{Degen}},
  \bibinfo{author}{\bibfnamefont{H.~J.} \bibnamefont{Mamin}}, \bibnamefont{and}
  \bibinfo{author}{\bibfnamefont{D.}~\bibnamefont{Rugar}},
  \bibinfo{journal}{Phys. Rev. Lett.} \textbf{\bibinfo{volume}{99}},
  \bibinfo{pages}{017201} (\bibinfo{year}{2007}).

\bibitem[{\citenamefont{{H{\"o}hberger Metzger} and Karrai}(2004)}]{metzger04}
\bibinfo{author}{\bibfnamefont{C.}~\bibnamefont{{H{\"o}hberger Metzger}}}
  \bibnamefont{and} \bibinfo{author}{\bibfnamefont{K.}~\bibnamefont{Karrai}},
  \bibinfo{journal}{Nature (London)} \textbf{\bibinfo{volume}{432}},
  \bibinfo{pages}{1002} (\bibinfo{year}{2004}).

\bibitem[{\citenamefont{Harris et~al.}(2007)\citenamefont{Harris, Zwickl, and
  Jayich}}]{harris07}
\bibinfo{author}{\bibfnamefont{J.~G.~E.} \bibnamefont{Harris}},
  \bibinfo{author}{\bibfnamefont{B.~M.} \bibnamefont{Zwickl}},
  \bibnamefont{and} \bibinfo{author}{\bibfnamefont{A.~M.}
  \bibnamefont{Jayich}}, \bibinfo{journal}{Rev. Sci. Instrum.}
  \textbf{\bibinfo{volume}{78}}, \bibinfo{pages}{013107}
  (\bibinfo{year}{2007}).

\bibitem[{\citenamefont{Arcizet~\emph{et
  al}.}(2006{\natexlab{b}})}]{arcizet_nov06}
\bibinfo{author}{\bibfnamefont{O.}~\bibnamefont{Arcizet~\emph{et al}.}},
  \bibinfo{journal}{Nature (London)} \textbf{\bibinfo{volume}{444}},
  \bibinfo{pages}{71} (\bibinfo{year}{2006}{\natexlab{b}}).

\bibitem[{\citenamefont{Gigan~\emph{et al}.}(2006)}]{gigan06}
\bibinfo{author}{\bibfnamefont{S.}~\bibnamefont{Gigan~\emph{et al}.}},
  \bibinfo{journal}{Nature (London)} \textbf{\bibinfo{volume}{444}},
  \bibinfo{pages}{67} (\bibinfo{year}{2006}).

\bibitem[{\citenamefont{Corbitt~\emph{et al}.}(2007)}]{corbitt07}
\bibinfo{author}{\bibfnamefont{T.}~\bibnamefont{Corbitt~\emph{et al}.}},
  \bibinfo{journal}{Phys. Rev. Lett.} \textbf{\bibinfo{volume}{98}},
  \bibinfo{pages}{150802} (\bibinfo{year}{2007}).

\bibitem[{\citenamefont{Schliesser~\emph{et al}.}(2006)}]{schliesser06}
\bibinfo{author}{\bibfnamefont{A.}~\bibnamefont{Schliesser~\emph{et al}.}},
  \bibinfo{journal}{Phys. Rev. Lett.} \textbf{\bibinfo{volume}{97}},
  \bibinfo{pages}{243905} (\bibinfo{year}{2006}).

\bibitem[{\citenamefont{Milatz et~al.}(1953)\citenamefont{Milatz, Van~Zolingen,
  and Van~Iperen}}]{milatz53}
\bibinfo{author}{\bibfnamefont{J.~M.~W.} \bibnamefont{Milatz}},
  \bibinfo{author}{\bibfnamefont{J.~J.} \bibnamefont{Van~Zolingen}},
  \bibnamefont{and} \bibinfo{author}{\bibfnamefont{B.~B.}
  \bibnamefont{Van~Iperen}}, \bibinfo{journal}{Physica (Amsterdam)}
  \textbf{\bibinfo{volume}{19}}, \bibinfo{pages}{195} (\bibinfo{year}{1953}).

\bibitem[{\citenamefont{Wineland~\emph{et al}.}()}]{wineland06}
\bibinfo{author}{\bibfnamefont{D.~J.} \bibnamefont{Wineland~\emph{et al}.}},
  \eprint{arXiv:quant-ph/0606180}.

\bibitem[{\citenamefont{Butt and Jaschke}(1995)}]{butt95}
\bibinfo{author}{\bibfnamefont{H.-J.} \bibnamefont{Butt}} \bibnamefont{and}
  \bibinfo{author}{\bibfnamefont{M.}~\bibnamefont{Jaschke}},
  \bibinfo{journal}{Nanotechnology} \textbf{\bibinfo{volume}{6}},
  \bibinfo{pages}{1} (\bibinfo{year}{1995}).

\bibitem[{d0x()}]{d0x}
\bibinfo{note}{This expression for $d$ neglects the curvature of the flexural
  mode and is strictly true only for $h \ll h_\mathrm{c}$ and $d_0 \ll w,\ h$.}

\bibitem[{\citenamefont{Marquardt et~al.}(2007)\citenamefont{Marquardt, Chen,
  Clerk, and Girvin}}]{marquardt07}
\bibinfo{author}{\bibfnamefont{F.}~\bibnamefont{Marquardt}},
  \bibinfo{author}{\bibfnamefont{J.~P.} \bibnamefont{Chen}},
  \bibinfo{author}{\bibfnamefont{A.~A.} \bibnamefont{Clerk}}, \bibnamefont{and}
  \bibinfo{author}{\bibfnamefont{S.~M.} \bibnamefont{Girvin}},
  \bibinfo{journal}{Phys. Rev. Lett.} \textbf{\bibinfo{volume}{99}},
  \bibinfo{pages}{093902} (\bibinfo{year}{2007}).

\bibitem[{\citenamefont{Wineland and Dehmelt}(1975)}]{wineland75}
\bibinfo{author}{\bibfnamefont{D.~J.} \bibnamefont{Wineland}} \bibnamefont{and}
  \bibinfo{author}{\bibfnamefont{H.~G.} \bibnamefont{Dehmelt}},
  \bibinfo{journal}{J. Appl. Phys.} \textbf{\bibinfo{volume}{46}},
  \bibinfo{pages}{919} (\bibinfo{year}{1975}).

\bibitem[{unc()}]{uncertainties}
\bibinfo{note}{The uncertainties for these numbers are a few micrometers,
  stemming from spatial nonuniformity in our etch process.}

\bibitem[{\citenamefont{Wilson-Rae et~al.}(2007)\citenamefont{Wilson-Rae,
  Nooshi, Zwerger, and Kippenberg}}]{wilson-rae07}
\bibinfo{author}{\bibfnamefont{I.}~\bibnamefont{Wilson-Rae}},
  \bibinfo{author}{\bibfnamefont{N.}~\bibnamefont{Nooshi}},
  \bibinfo{author}{\bibfnamefont{W.}~\bibnamefont{Zwerger}}, \bibnamefont{and}
  \bibinfo{author}{\bibfnamefont{T.~J.} \bibnamefont{Kippenberg}},
  \bibinfo{journal}{Phys. Rev. Lett.} \textbf{\bibinfo{volume}{99}},
  \bibinfo{pages}{093901} (\bibinfo{year}{2007}).

\end{thebibliography}

\end{document}